# Atomic Structure of Grain Boundaries, Dislocations and Associated Strain in Templated Co-evaporated Photoactive Halide Perovskites

Huyen T Pham[1], Siyu Yan[2], Zhou Xu[3], Weilun Li[1], Sergey Gorelick[4], Michael B Johnston[2]*, Joanne Etheridge[1, 3]*

[1]School of Physics and Astronomy, Monash University, Wellington Road, Clayton, 3800, Victoria, Australia
[2]Department of Physics, University of Oxford, Clarendon Laboratory, Parks Road, Oxford, OX1 3PU, United Kingdom.
[3]Monash Centre for Electron Microscopy, Monash University, Wellington Road, Clayton, 3800, Victoria, Australia.
[4]Ramaciotti Centre for Cryo-Electron Microscopy, Monash University, Wellington Road, Clayton, 3800, Victoria, Australia.

Email Address: joanne.etheridge@monash.edu

**Abstract.** Structural defects, particularly grain boundaries, play a crucial role governing charge transport and the optoelectronic properties of metal halide perovskites, thereby limiting the performance of devices. Solar cells incorporating templated $FA_{0.9}Cs_{0.1}PbI_{3-x}Cl_x$ show significant improvements in grain orientation and steady-state power conversion efficiency, however, the underlying mechanisms remain unclear. In this study, we address this gap by employing a suite of tailored low-dose electron microscopy techniques to investigate the templated $FA_{0.9}Cs_{0.1}PbI_{3-x}Cl_x$ film, revealing that it exhibits a preferred crystallographic orientation along the ⟨001⟩ zone axis, with arbitrary grain rotations about that axis, indicative of a Volmer–Weber growth mechanism. We determine the atomic structure of the resulting high-angle and low-angle grain boundaries. We also reveal the presence of edge dislocations and their associated strain fields, demonstrating the compressive strain on one side of the dislocation core and tensile strain on the opposite side. Furthermore, we find dislocations associated with stacking faults. These atomic-level insights uncover which grain boundaries and intra-grain defects are likely to act as recombination centres or modify bandgaps, crucial for understanding which defects influence the performance of perovskite solar cell devices.

# 1. Introduction

Over the last two decades, organic-inorganic metal halide perovskite (OIMHP) semiconductors have become one of the most promising photovoltaic materials for next-generation solar cells. They can be used as the main absorber layer in single-junction perovskite solar cells (PSCs) or combined with silicon in tandem configuration solar cells [1]. PSC technology has improved rapidly, with power conversion efficiency (PCE) increasing from 3.8 % in 2009 to the current record of 27 % [2] for single-junction PSCs. However, the optimisation of these photovoltaic devices faces several challenges, including current-voltage hysteresis, toxicity of lead, ion migration, and vulnerability to degradation when exposed to light, moisture, heat, air, and electron beams [3]. Global efforts have advanced mixed-halide perovskites through alloying inorganic cations (Cs, Rb), organic cations (MA, FA), and halides (I, Br, Cl) [4–6].

The addition of chlorine anions (Cl$^-$) to the perovskite precursor has been shown to improve the morphology, crystallinity, grain size and optoelectronic properties of OIMHPs materials, as well as the power conversion efficiency (PCE) and long-term stability of PSC devices [7–11]. In 2023, PSC devices using alkylammonium chlorides (RACl) added to FAPbI$_3$ have reached a record of 26.08 % (certified 25.73 %) efficiency under standard illumination [11]. As of early 2026, PSCs have achieved 34.85% efficiency in FA$_{1-x}$Cs$_x$Pb(I$_{1-y}$Br$_y$)$_3$ perovskite/silicon tandem devices (LONGi Solar, verified by National Laboratory of the Rockies) and 27.87% in single-junction cells (SolaEon, certified by China's National Photovoltaic Industry Metrology). However, the underlying mechanisms of how Cl$^-$ anions and Cs$^+$ cations affect the microstructure, crystal structure and defects and improve the efficiency and stability of the perovskite photo-absorber are still unclear.

Most OIMHP materials are polycrystalline and can have defects at the boundaries where two grains meet, as the structure tries to accommodate crystallographic misalignment between grains. Grain boundaries (GBs) in OIMHP films can introduce defect states and accumulate impurities, which can have a significant impact on the performance and stability of solar cell devices [12–14]. For example, grain boundaries (GBs) can incur a range of defects, including points defects such as Pb-I anti-site defects and I$^-$/Br$^-$ or MA$^+$/FA$^+$ vacancies, under-coordinated X$^-$ or Pb$^{2+}$ ions, lead clusters and phase segregation [12,15–17]. Efforts to minimize these issues are key to advancing the efficiency and stability of perovskite-based devices [16,17].

In OIMHP materials, GBs have been studied using scanning/transmission electron microscopy (S/TEM), scanning electron microscopy (SEM), solid-state electron backscatter diffraction (EBSD), Kelvin probe microscope, atomic force microscopy, and photoluminescence microscopy [12]. Low-dose STEM has been used to study the morphology and atomic structure of OIMHP materials [18–20]. Both low-angle GBs and high-angle GBs have been observed [12,18]. However, there are relatively few studies of the atomic structure at the GBs due to their delicate structure, which can be easily modified under electron beam exposure, particularly at defect sites, such as GBs [21].

In addition to GBs, intragrain defects are important. In conventional photovoltaic materials such as silicon and CdTe, intra-grain defects, including stacking faults, dislocations, and twins, are major sources of charge-carrier recombination [22,23]. Planar defects (e.g., twins, stacking faults, and Ruddleson–Popper faults) have also been identified in OIMHP solar cells [7,13,18,24–32]. Depending on their specific atomic-structure, they can strongly affect charge-carrier mobility, ion migration, defect states, and overall device efficiency, for instance, in $FAPbI_3$, a {111} stacking fault is structurally equivalent to a monolayer of the photo-inactive "delta" phase [7,13,18,24–32]. Stacking faults can also be a nucleation site for dislocations and introduce deep-level states in the bandgap, enhancing charge-carrier recombination and reducing solar cell efficiency [33,34]. While significant progress has been made, there is still much to understand about the atomic structure of both GBs and intragrain defects and their impact on PSC properties.

Recently, PSC devices using a templating $FA_{0.9}Cs_{0.1}PbI_{3-x}Cl_x$ layer demonstrated significant enhancements in grain orientation and improved device performance [35]. Introducing an ultrathin (15 nm) templating layer consisting of sequentially-grown stoichiometric OIMHP between the perovskite layer and the substrate produced highly oriented co-evaporated perovskite films with consistent morphology, structure, and optoelectronic properties across various substrate materials. X-ray diffraction (XRD) analysis reveals that the $FA_{0.9}Cs_{0.1}PbI_{3-x}Cl_x$ perovskite film grown on a templating layer exhibits enhanced crystallinity and a higher degree of preferred orientation. Grazing incidence wide-angle X-ray scattering further confirms the improved crystallographic alignment of the templated $FA_{0.9}Cs_{0.1}PbI_{3-x}Cl_x$ films, as evidenced by the presence of discrete maxima in the diffraction pattern, showing a more highly ordered crystalline structure. However, XRD is primarily a volume-based method and does not provide details about the relative orientation of adjacent grains, and the atomic structure at GBs.

In this work, we develop and apply ultralow electron dose microscopy methods to examine the relative orientation of grains, GB structure and associated strain and contaminant phases in co-evaporated $FA_{0.9}Cs_{0.1}PbI_{3-x}Cl_x$ photoactive perovskite, uncovering how the growth process governs boundary structure and defect formation at the atomic scale. This was accomplished by utilizing a four-dimensional (4D)-STEM capability built within an SEM to allow ultralow-dose, large area grain orientation mapping, together with ultralow-dose high resolution STEM in a TEM to characterise reliably the atomic structure at GBs, intragrain defects, and local strain. These results provide new atomic-scale insights into the growth mechanism and structural origins of defect formation and their implications for PSC performance.

## 2. Results

## 2.1. Grain orientation and crystallographic phase: 4D-STEM in SEM

A popular method for mapping grain orientation is EBSD [36,37]. However, EBSD requires a relatively high electron dose to generate sufficient contrast for indexing [38] because it uses focused electron probes with a high electron current density and EBSD patterns are formed by the diffraction of only a small fraction of inelastically scattered electrons. The overall signal yield is inherently limited. OIMHPs exhibit pronounced sensitivity to electron irradiation, readily undergoing structural degradation, ion migration, and radiolysis under conventional electron microscopy conditions, thereby strongly motivating the development and adoption of low-dose methodologies in place of conventional EBSD techniques [18,19,39,40].

As a lower dose alternative, we use a 4D-STEM configuration that we have built within an SEM, as shown in Figure 1a (the experimental set up is described in Supporting Information, S2). We scan a nanometre-diameter (2 nm), approximately collimated electron probe across the specimen and at each position of the probe, we record a diffraction pattern (comprising both elastic and inelastically scattered electrons). From this we can determine crystallographic information, including the crystal phase and orientation (as described in detail in Supporting Information, S2). Compared to TEM-based 4D-STEM, the SEM implementation offers a significantly larger field of view. The lower electron energy inherent to SEM mitigates knock-on damage, preserving the structural integrity of radiation-sensitive perovskite materials. Figure S2 (Supporting Information) shows the morphology and microstructure of the $FA_{0.9}Cs_{0.1}PbI_{3-x}Cl_x$ perovskite film, with an average grain size of approximately 98 nm estimated from over 500 grains in the low-magnification TEM image.

Using 4D-STEM within the SEM, we obtained 40,000 diffraction patterns spanning a 5 μm × 5 μm area of the specimen (recorded at 46 nm intervals to further minimize the impact of the electron beam, such as heating and knock-on damage). Figure 1b shows a subset of these electron diffraction patterns, each corresponding to a different electron probe scanning position. The diffraction patterns reveal that the majority of grains in an evaporated thin film of FA$_{0.9}$Cs$_{0.1}$PbI$_{3-x}$Cl$_x$ exhibit an average cubic crystal structure consistent with the $Pm\bar{3}m$ space group reported previously, and are predominantly oriented along the ⟨001⟩ zone axis, indicating a high degree of crystallographic alignment across the film.[35] Additionally, occasional regions of the non-photoactive PbI$_2$ phase are also detected. Specifically, we observed 89 diffraction patterns corresponding to the PbI$_2$ hexagonal phase out of 40,000 patterns, accounting for ~0.22 % of the diffraction patterns.

To further analyse this data, we reconstruct a dark-field STEM image from the 4D-STEM dataset using the py4D-STEM software package [41]. This approach allows the morphology and local structural variations within the OIMHP film to be visualised across a large area, albeit at low magnification (Fig. 1c).

Example diffraction patterns are shown in Figure 1d, generated from the red dot positions highlighted in Figure 1c, and correspond to the cubic perovskite phase, which is oriented along the ⟨001⟩ zone axis. In contrast, the diffraction patterns shown in Figure 1e, deriving from the bright blue probe positions, are consistent with the hexagonal PbI$_2$ phase with the orientation aligned along the ⟨$\bar{8}$81⟩ zone axis, indicating the structural difference between the two phases.

From the 4D-STEM results, we found that the templated FA$_{0.9}$Cs$_{0.1}$PbI$_{3-x}$Cl$_x$ film has crystal grains that are almost exclusively orientated with the ⟨001⟩ zone axis perpendicular to the substrate (as seen in Fig 1g) but are randomly rotated about this axis (as seen in Fig 1f). These results are consistent with the X-ray findings[35]. This is suggestive of a Volmer–Weber nucleation and growth process,[42] whereby each grain nucleates independently and grows perpendicular to the substrate in the ⟨001⟩ direction. As the grains widen, they "collide" with adjacent grains at arbitrary angles.

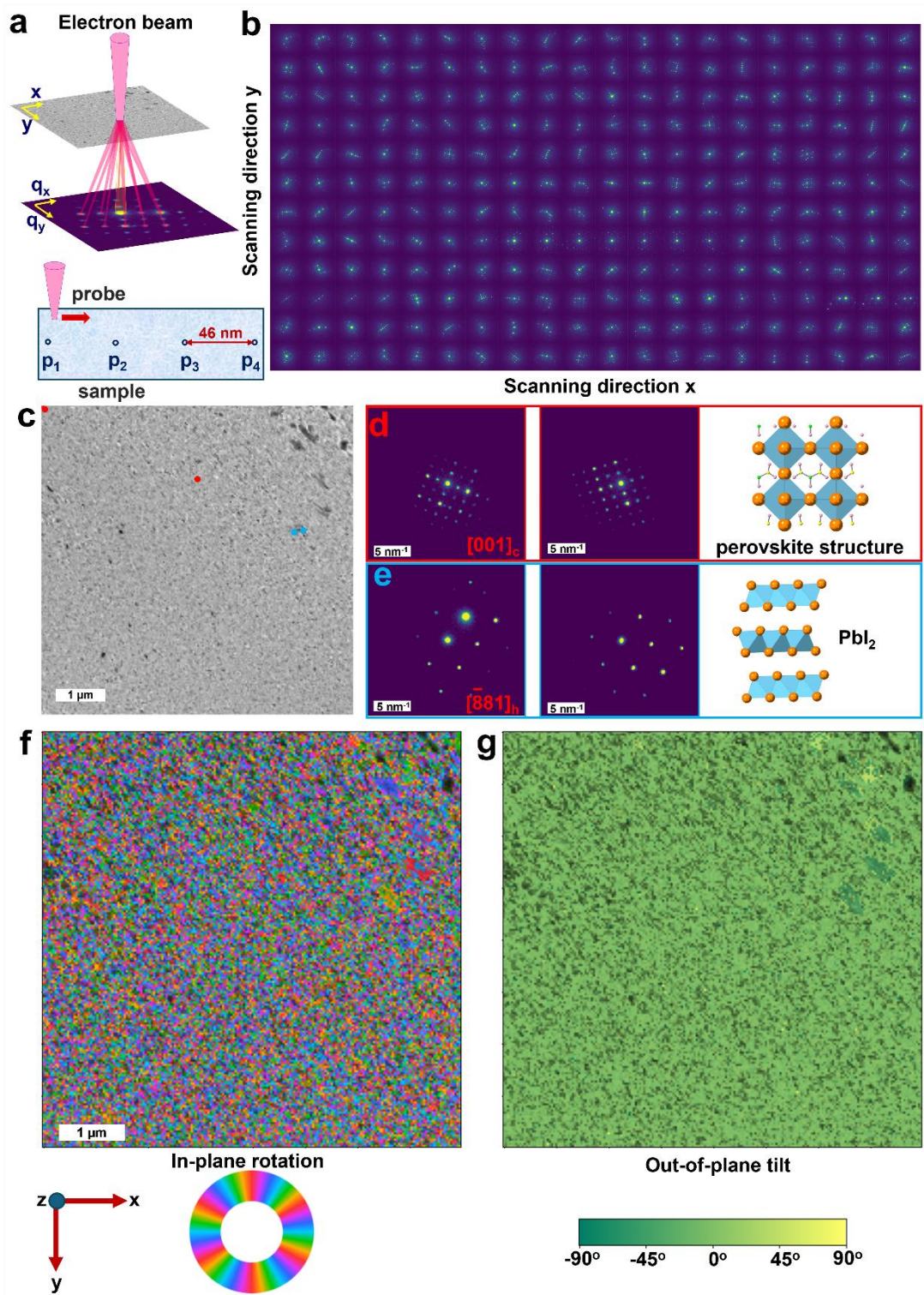

**Figure 1.** 4D-STEM of the $FA_{0.9}Cs_{0.1}PbI_{3-x}Cl_x$ film. (a) The geometry of the 4D-STEM experiment in the SEM. (b) Subset of the series of 40,000 diffraction patterns of the $FA_{0.9}Cs_{0.1}PbI_{3-x}Cl_x$ film. (c) Reconstructed dark-field image of the $FA_{0.9}Cs_{0.1}PbI_{3-x}Cl_x$ film from the 4D-STEM dataset. (d, e) Diffraction patterns generated from the positions indicated by the red and blue dots in (c), respectively. (f) In-plane grain orientation map along x and y axes, i.e. rotation around the ⟨001⟩ axis and (g) out-of-plane grain orientation map along z or ⟨001⟩ axis.

## 2.2. High-angle grain boundaries

High-angle GBs are the interfaces between two grains in polycrystalline materials, characterised by a significant difference in crystal orientation, typically greater than 15°. We characterised the atomic structure of the GBs between adjacent grains in detail, using low dose high-resolution low-angle annular dark field (LAADF) STEM to capture clear and precise images of the boundaries.

Figure S3 (Supporting Information) presents high-magnification, low-dose LAADF-STEM images of the cubic $FA_{0.9}Cs_{0.1}PbI_{3-x}Cl_x$ structure along the ⟨001⟩ zone axis, before and after Butterworth filtering. Butterworth-filtered LAADF STEM images (Fig. 2a, c, e, g, i and k) provide a clear representation of the GBs in the co-evaporated thin film of $FA_{0.9}Cs_{0.1}PbI_{3-x}Cl_x$, revealing several significant features. Firstly, each individual grain is well-aligned with the ⟨001⟩ zone axis perpendicular to the substrate, consistent with the 4D-STEM results. The grain interiors exhibit high crystallinity that extends right across the interior, indicating that the GBs do not disrupt the crystal quality of the interior perovskite material. Secondly, in most cases, adjacent grains have an arbitrary relative orientation, with incoherent interface structures, again consistent with the 4D-STEM results.

Importantly, there is typically no observable amorphous material at the GBs, which further supports the notion that these boundaries do not significantly contribute to structural disorder within the material. However, the incoherent interfaces observed in Figure 2 result in point defects and dangling bonds which are likely to act as harmful recombination centres.

The Fourier transforms (FT) of the Butterworth-filtered LAADF images (Fig. 2a, c, e, g, i, and k) are shown in Figures 2b, d, f, h, j, and l, respectively. These FT images confirm that each grain maintains a well-defined cubic structure, with its orientation aligned along the ⟨001⟩ zone axis. These FTs enable the relative rotation angle of adjacent grains to be determined. A range of high ($\theta$) angle GBs are observed, including 26°, 20.8°, 36.9°, 41.7°, and 28.7°, again consistent with 4D-STEM results.

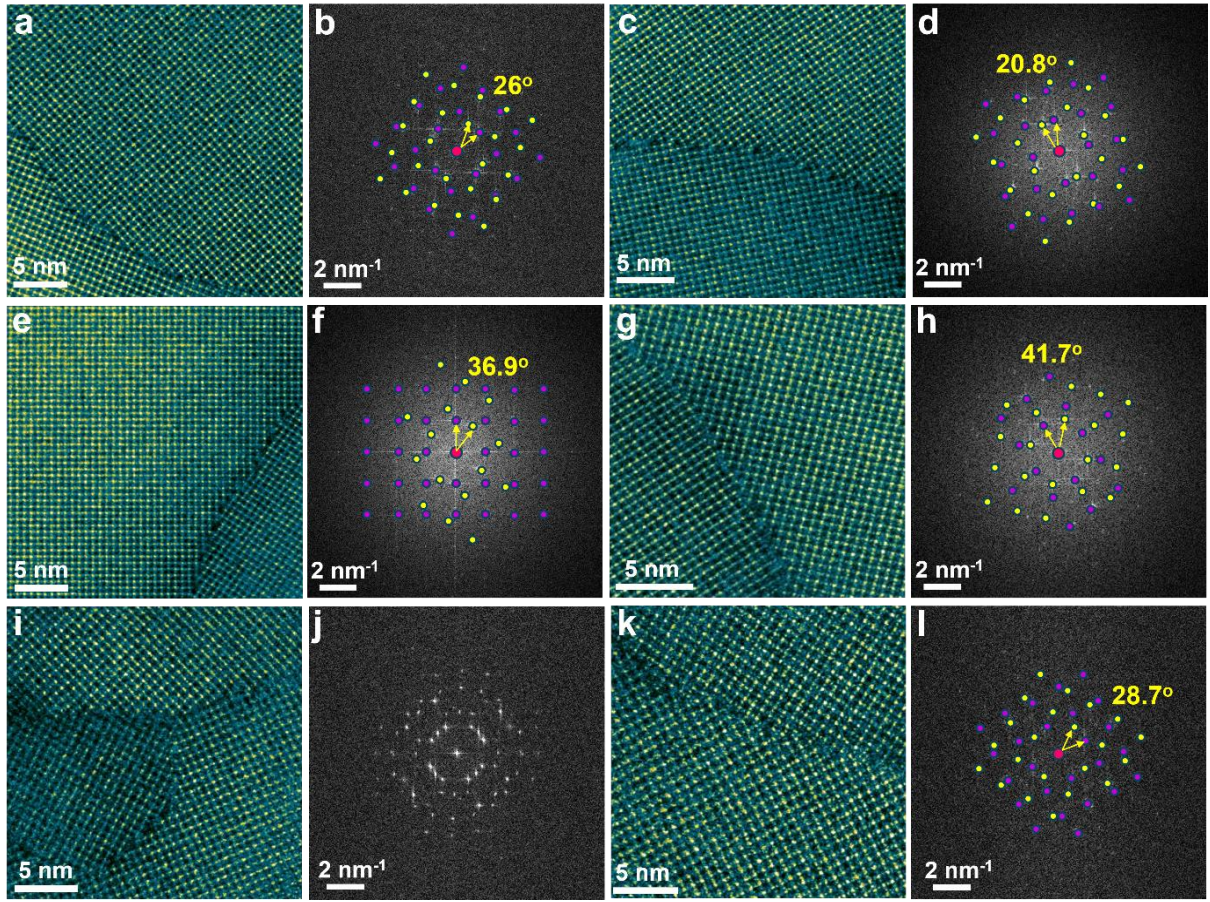

**Figure 2.** Atomic-resolution STEM images revealing the GBs in a $FA_{0.9}Cs_{0.1}PbI_{3-x}Cl_x$ thin film with a variety of GB angles. (a, c, e, g, i, k) Low-dose low-angle annular dark field (LAADF) STEM image of GBs of $FA_{0.9}Cs_{0.1}PbI_{3-x}Cl_x$ film is filtered with a Butterworth filter to enhance contrast. (b, d, f, h, j, l) The FT corresponding to the LAADF STEM images in (a, c, e, g, i, k), respectively, reveal the relative rotation angle about the ⟨001⟩ axis between adjacent grains.

Unlike the arbitrary high-angle grain boundary orientations in Figure 2, Figure 3a-b and Figure S4 (Supporting Information) show a coincidence site lattice (CSL) GB. In a CSL GB, the GB angle is such that certain lattice sites in one grain align along the GB with the corresponding sites in the adjacent grain, creating a set of matching lattice sites that are shared by both grains. These "coincident sites" appear regularly along the boundary, contributing to the CSL structure. Compared to random GBs, CSL GBs are thought to have lower energy because the atoms fit together more efficiently [43–45]. As a result, CSL GBs are crucial in materials science, as they can significantly impact the material's mechanical, electronic, and thermal properties.

The Butterworth-filtered LAADF image shows a CSL GB that is atomically flat and coherent with no steps, secondary phases, intermediate layers or amorphous layers, indicating a clean and direct bonding of the two single crystals across the GB. The GB displays a mirror symmetry in this projection, however, not all atoms in the bulk lattice meet at the GB, just a specific subset of coincident atomic sites.

A Σ5 (310)/[001] GB is a CSL boundary in which one in five lattice sites coincides between adjacent crystal grains. It is a symmetric tilt boundary about the ⟨001⟩ axis and exhibits mirror symmetry in this projection. The misorientation angle ($\theta$) is 36.9°, corresponding to twice the angle (18.45°) between the (310) and (100) planes, and is consistent with the Σ5 relationship, as confirmed by FT and simulated diffraction patterns (Fig. 3c). The boundary structure consists of periodically arranged structural units (commonly described as kite-shaped units), within which coincidence sites contribute to the relatively low grain boundary energy.

The atomic structure models for grain A and grain B are presented in Figures 3d and 3e, respectively. Both grains A and B are observed along the ⟨001⟩ zone axis, with grain B rotated by 36.9° relative to grain A. The resulting atomic configuration at the Σ5 (310)/[001] GB is shown in Figure 1f. This special type of boundary may significantly influence the material's properties, such as its mechanical strength and electrical conductivity [46].

In Figures 3g–i, we present the simulated electron diffraction patterns corresponding to grain A, grain B, and the Σ5 (310)/[001] GB, consistent with the experimental results displayed in Figure 3c, and confirming the rotation angle between two grains at Σ5 (310)/[001] GB.

Figure S5 (Supporting Information) schematically presents a superimposed atomic model of the Σ5 (310)/[001] GB, highlighting a precise rotation relationship of $\theta = 36.9°$. For clarity and simplicity, the positions of the A-site cations are omitted in this figure. The black arrow indicates the virtual GB plane, providing a reference for the geometric relationship between the grains. Upon closer inspection, numerous coincident sites become visible, where iodine atoms from two adjacent grains overlap, as highlighted by the yellow circles. These coincident iodine atoms form a CSL, which plays a crucial role in determining the structural properties and stability of the GB.

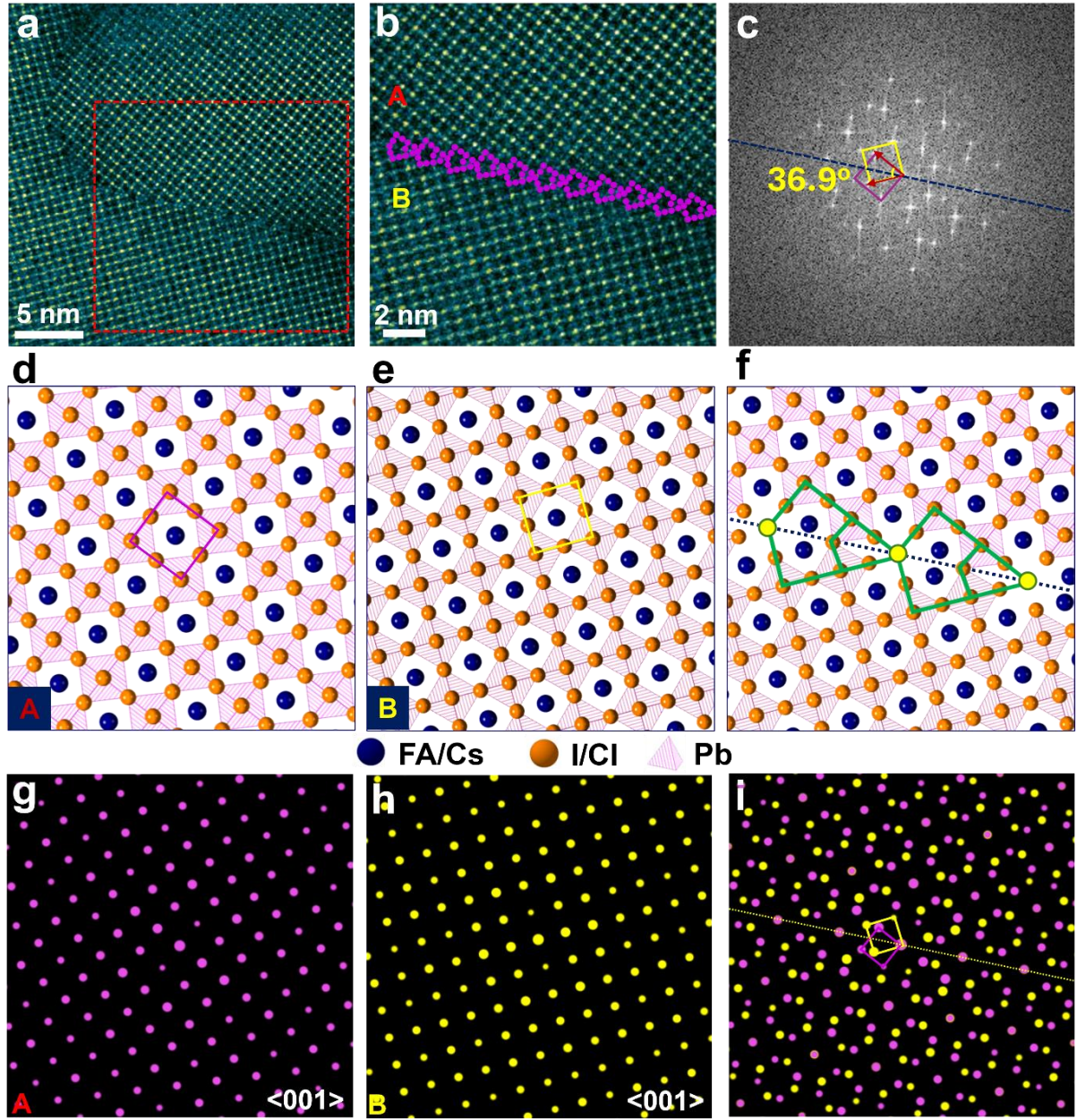

**Figure 3.** Σ5 (310)/[001] GBs in FA$_{0.9}$Cs$_{0.1}$PbI$_{3-x}$Cl$_x$ film. (a) Low-dose LAADF-STEM image showing atomic resolution detail of the Σ5 (310)/[001] GBs, a Butterworth filter has been applied to enhance contrast. (b) Zoomed-in low-dose LAADF-STEM image of the region marked by the red square in Figure a, showing the atomic structure at the boundary of the Σ5 (310)/[001] GBs. (c) The FTs corresponding to the LAADF image in (a) with simulated electron diffraction pattern of Σ5 (310)/[001] GBs. (d, e) Schematic of the ⟨001⟩ projected atomic structure of grain A and grain B. (f) Schematic of the ⟨001⟩ projected atomic structure of Σ5 (310)/[001] GBs. (g, h) Simulated electron diffraction of grain A and grain B. (i) Simulated electron diffraction of Σ5 (310)/[001] GBs.

## 2.3. Low-angle grain boundaries and edge dislocations

A low-angle GB in a polycrystalline material is where two grains meet with a small relative rotation angle. We also observe low-angle GBs in the co-evaporated thin film of $FA_{0.9}Cs_{0.1}PbI_{3-x}Cl_x$ where the crystallographic misorientation angle between the grains is less than 15° (Figure 4 and Figures S6, Supporting Information). Furthermore, we observe the presence of edge dislocations associated with these low-angle boundaries. These dislocations are significant because they can influence both the mechanical deformation behaviour and the charge transport properties of the material [47,48].

Figure 4a shows a typical example where the misorientation angle between two grains is approximately 5.3° (Fig. 4c). We observed two types of edge dislocations associated with this low-angle boundary, as shown in Figure 4b, highlighted as numbers 1 and 2. The first edge dislocation structure (highlighted as number 1 in Fig. 4b) can be understood as the result of removing half of a {010} plane from the perfect lattice, creating interacting dangling bonds that extend into the core region (Fig. 4d). This dislocation core structure has been previously observed in diamond and other crystalline materials[49–51], where it plays an important role in influencing the mechanical properties of the material.

The second edge-type dislocation is labelled as number 2 in Figure 4b. This dislocation forms by shifting the core of the first dislocation along the glide plane, causing the core to alternate back and forth along the dislocation line (Fig. 4e), leaving dangling bonds. These dangling bonds are energetically unfavourable, leading to localized electronic states that are likely to form recombination centres.

The strain field surrounding the dislocations is mapped and analysed using geometrical phase analysis (GPA), which enables atomic-scale visualization of strain in STEM [52]. This approach enables the detection of variations in inter-atomic distances, offering valuable insights into the local deformation around dislocations. The GPA results corresponding to the edge dislocation are shown in Figure 4f-h. (To ensure that the results accurately reflect the true strain field, the GPA analysis was performed on the raw experimental data, without any pre-processing through Butterworth filtering).

In Figure 4f, the strain is defined relative to the bulk region, located sufficiently far from the dislocation core, where it is assumed that the strain is effectively zero. The strain components, including $\varepsilon_{xx}$ (strain component along the $x$-direction ({001})), $\varepsilon_{xy}$ (shear strain), and $\varepsilon_{yy}$ (strain component along the $y$-direction ({001})) and the magnitude of strain is shown in the colour scale. In the $\varepsilon_{xx}$, $\varepsilon_{yy}$, and $\varepsilon_{xy}$ maps, blue indicates a negative value, while pink corresponds to a positive value.

A single edge dislocation in the OIMHP thin film is characterised by a strain distribution that is consistent with compressive strain on one side of the dislocation core and a tensile strain on the opposite side.

From the $\varepsilon_{xx}$ and $\varepsilon_{yy}$ visualization in Figure 4f, h, it is evident that the perovskite lattice experiences a compressive strain (negative) on the side of the dislocation associated with the extra {010} halfplane. This compression occurs as the lattice is locally pushed together by the dislocation's presence. Conversely, on the opposite side of the dislocation, the lattice undergoes tensile stress (positive), where the atomic spacing is increased due to the dislocation's core structure. These distinct strain patterns are characteristic of the edge dislocation configuration. The strain maps also highlight the presence of several dislocations along the boundary.

Another example of a low-angle GB and associated dislocations is presented in Figures S6 (Supporting Information). This GB has a misorientation angle of 7° between the two grains and induces several edge dislocations, as highlighted in Figures S6a, b (Supporting Information). The strain distributions of the edge dislocations, including $\varepsilon_{xx}$, $\varepsilon_{xy}$, $\varepsilon_{yy}$ are shown in Figures S6d-f (Supporting Information). These findings are consistent with the observations above. Specifically, we identify five dislocation cores across the low-angle GB. The strain components reach their maximum at the core, transitioning from positive to negative strain.

A further example of a low angle (4°) GB is given in Figures S7 (Supporting Information). The grain contains both edge dislocations and stacking faults, which are highlighted in Figure S7b with behaviour consistent with the dislocations described above.

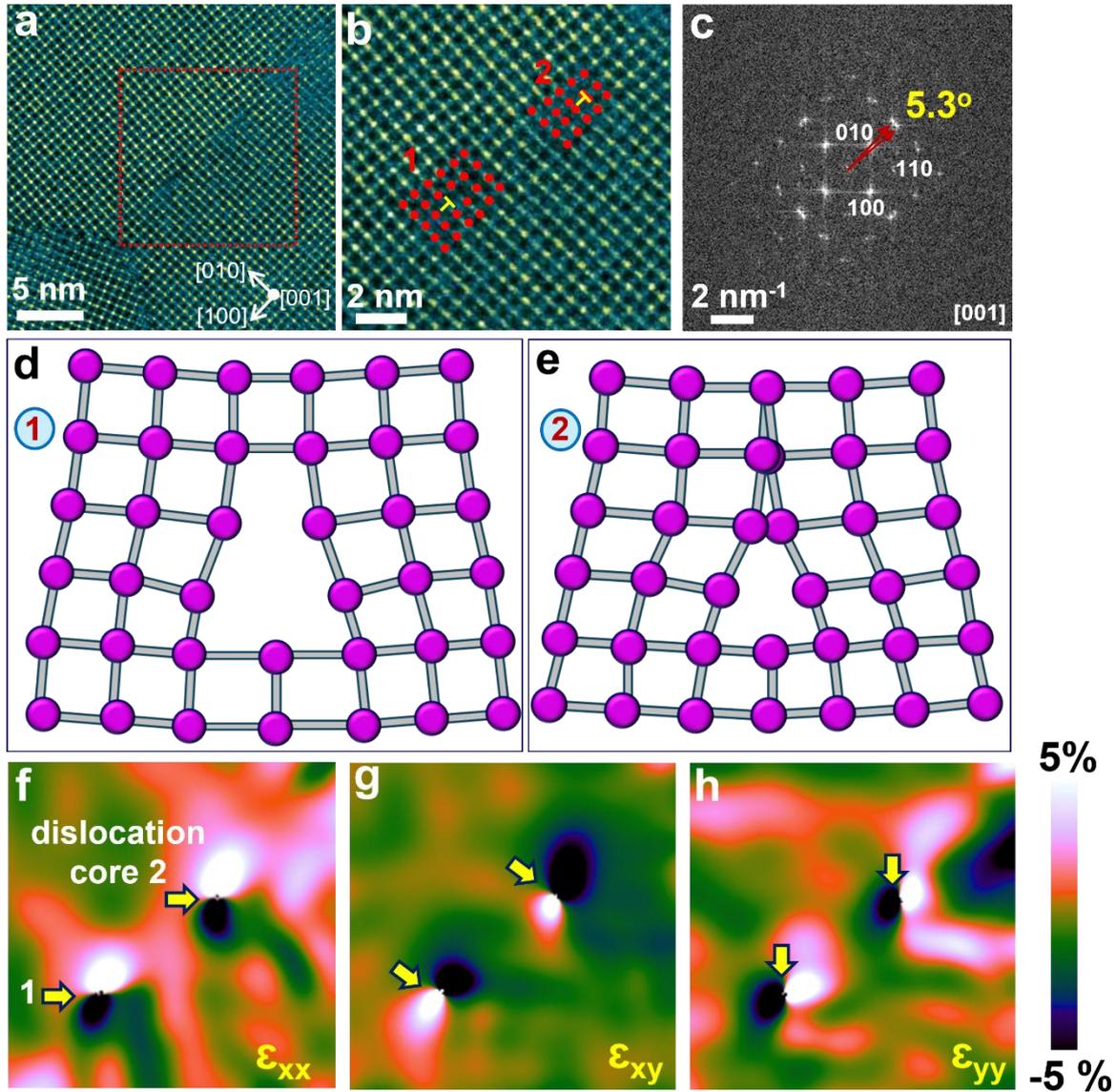

**Figure 4.** Low-angle GB and associated edge dislocations in $FA_{0.9}Cs_{0.1}PbI_{3-x}Cl_x$ film. (a) Low-dose atomic-resolution LAADF-STEM image of a low-angle grain boundary, with a misorientation angle of 5°. The image is filtered with a Butterworth filter to enhance contrast. (b) Zoomed-in low-dose LAADF-STEM image of the region marked by the red square in (a), showing the atomic structure of the edge dislocation (c) FTs of the low-dose LAADF-STEM image shown in (a). (d, e) Atomic structures of the two different edge dislocation cores. (f-h) Strain distributions $\varepsilon_{xx}$, $\varepsilon_{xy}$, $\varepsilon_{yy}$ in the image from (b), generated using GPA analysis, showing strain fields surrounding the two dislocation cores. The perpendicular symbol (⊥) indicates the edge dislocation.

## 2.4. 90-degree planar GBs (or pseudo-twin boundaries) and associated stacking fault and dislocations.

Intra-grain planar defects, such as {111} stacking faults (or equivalently {111} twin boundaries when in isolation), are commonly observed in FA-based PSCs, and have been shown to be detrimental to device performance [24,53–55]. These {111} stacking faults or twin boundaries involve face-sharing octahedra. In the $FA_{0.9}Cs_{0.1}PbI_{3-x}Cl_x$ films, we observe a different type of 'pseudo-twin' boundary, parallel to the {110} plane and involving edge-sharing octahedra.

Figure 5a shows a Butterworth filtered LAADF-STEM image of the co-evaporated $FA_{0.9}Cs_{0.1}PbI_{3-x}Cl_x$ film along the ⟨001⟩ direction. The regular atomic arrangement suggests a well-ordered crystal lattice, but a "zig-zag" boundary in the image and associated diffuse scattering in the FT of the image indicate the presence of planar defects parallel to the {110} plane. These planar defects could be thought of as a coherent 90° GB or a local {110} twin plane, as the structure is mirrored across the plane. However, the limited extent of each arm of the 'zig-zag' does not meet the exact criteria for a twin boundary (hence we nick-name it "pseudo-twin" or more formally, an "antiphase boundary").

An antiphase boundary in cubic perovskites arises when one region of the lattice is rotated by 90° relative to the adjoining region and displaced by ½{110} along the boundary, thereby generating a localized structural mismatch. As demonstrated by Chen *et al.* through density functional theory calculations, such an antiphase boundary preserves the overall cubic symmetry while introducing a localized defect that modifies the electronic and ionic properties [20].

These {110} planar defects, typically measuring between 5 and 10 nm in length, are not uniformly distributed throughout the thin film. This non-uniform distribution suggests that specific microstructural conditions, such as variations in local strain, composition, or growth dynamics, may favour the formation of these faults in some areas more than others. Two pseudo-twin planes are often seen to intersect at a 90° angle, forming a dislocation core at the intersection. We observed two dislocation cores in Figure 5b highlighted in number 1 and 2.

This boundary also results in an effective stacking fault parallel to the {100} planes. Atomic-scale analysis of the LAADF-STEM images (Fig. 5a, b) show that these stacking faults connect a row of Pb-I columns with a row of I⁻ columns along the ⟨001⟩ direction, (instead of a continuous row of Pb-I columns) (Fig. 5b). In this arrangement, the $[PbI_6]^{4-}$ octahedra are edge-sharing parallel to the {110} plane, as opposed to corner-sharing, as shown in Figure 5d. When two octahedra share an edge, the two corners of one octahedron are aligned with two corners of the adjacent octahedron.

The GPA strain maps corresponding to $\varepsilon_{xx}$ (strain component along the x-direction), $\varepsilon_{yy}$ (strain component along the y-direction), and $\varepsilon_{xy}$ (shear strain) derived from the LAADF-STEM image presented in Figure 5a are shown in Figures 5e, g, and f, respectively. The stacking fault disrupts the local lattice periodicity, inducing significant in-plane strain in both the x and y directions, as illustrated in Figure 5e. The stacking fault has collective displacements parallel to the stacking fault plane and thus has the same strain status.

The strain component $\varepsilon_{xx}$ displays a distinct variation from one dislocation core to the next. Specifically, across core 1, $\varepsilon_{xx}$ transitions from a positive to a negative value, whereas across core 2, $\varepsilon_{xx}$ shifts from negative to positive, as clearly depicted in Figure 5e. This reversal of the strain field from one core to the next likely acts to reduce the overall lattice strain.

The shear strain $\varepsilon_{xy}$ exhibits similar features in its variation across the dislocation cores, as shown in Figure 5f. Specifically, $\varepsilon_{xy}$ transitions from a negative to a positive value across core 1, whereas $\varepsilon_{xy}$ shifts from positive to negative across core 2. The strain component $\varepsilon_{yy}$ shows a negative strain without $\varepsilon_{xx}$ reversion across both core 1 and core 2 due to constrained relaxation perpendicular to the defect plane.

In contrast, a different set of stacking faults was observed in the $FA_{0.9}Cs_{0.1}PbI_{3-x}Cl_x$ film, as shown in the Butterworth-filtered LAADF images, and the FT in Figure S8a-c (Supporting Information). In this case, the distribution of strain component $\varepsilon_{xx}$ and shear strain $\varepsilon_{xy}$ along the stacking fault is uniform, suggesting a consistent strain field across the fault line (Fig.S8 d-f). Notably, no significant variations were detected across the dislocation cores 3 and 4, indicating that these cores do not contribute to localized strain fluctuations.

The uniformity of strain across the fault and lack of variation near the dislocation cores implies that the atomic configuration at the cores plays a critical role in maintaining strain stability and inverting the strain distribution. These findings further emphasize the importance of understanding the atomic scale structure of dislocation cores for the design and optimization of materials with tailored mechanical properties, particularly in OIMHP films.

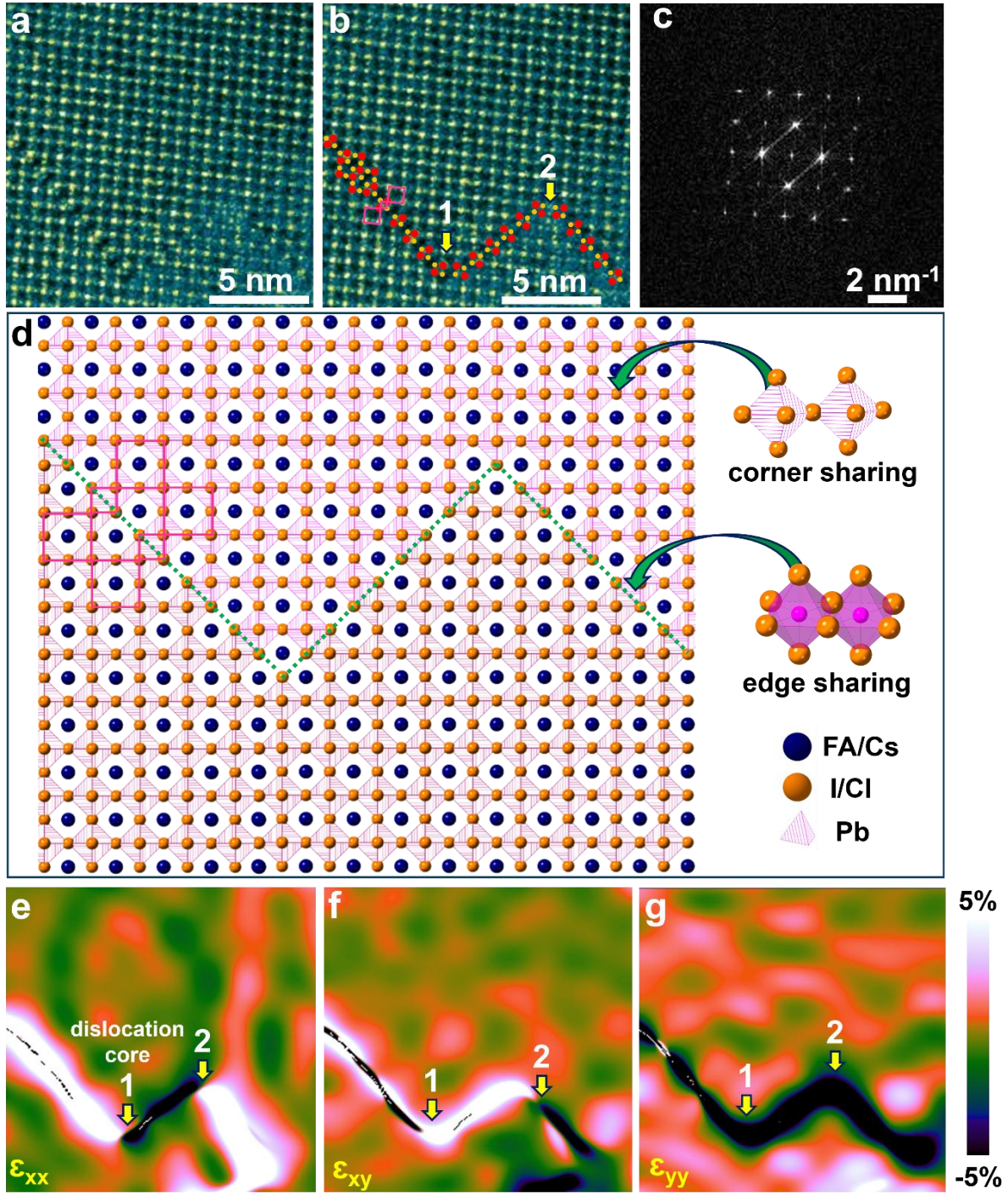

**Figure 5.** 90-degree {110} planar GBs (or pseudo-twin boundaries) and associated {100} stacking faulty and dislocations in the $FA_{0.9}Cs_{0.1}PbI_{3-x}Cl_x$ film. (a) Low-dose high-resolution LAADF-STEM image of the zig-zag 90-degree planar GB in cubic $FA_{0.9}Cs_{0.1}PbI_{3-x}Cl_x$ grains along the ⟨001⟩ zone axis. The image is filtered with a Butterworth filter to enhance contrast. (b) LAADF-STEM image with 90-degree {110} planar GB highlighted in $FA_{0.9}Cs_{0.1}PbI_{3-x}Cl_x$ film. (c) Fourier transform of the low-dose LAADF-STEM image shown in (a). (d) Schematic of the atomic structure of the planar defect. The corner-sharing octahedra become edge-sharing along the defect plane (e-g) Strain distributions $\varepsilon_{xx}$, $\varepsilon_{xy}$, $\varepsilon_{yy}$ in the image from (a), generated using GPA analysis. The dislocation cores are labelled with coloured arrows.

## 2.5. Perovskite/PbI$_2$ interface and associated dislocations

The presence of the PbI$_2$ phase has been consistently reported across a range of OIMHP compositions. Notably, low concentrations of PbI$_2$ do not appear to adversely impact photovoltaic performance; however, the underlying mechanisms responsible for this seemingly benign effect remain a subject of ongoing investigation [18,56].

In the 4D-STEM dataset discussed above (Figure 1), occasional PbI$_2$ phase was identified within the perovskite thin film in ~0.22% of diffraction patterns. The most frequently observed form of PbI$_2$ is the hexagonal 2H structure, whose characteristic reflections serve as a standard reference for identifying PbI$_2$ in perovskite films via XRD. However, previous studies have reported the presence of a minor fraction of the 3R trigonal PbI$_2$ phase, which is particularly significant due to its potentially coherent epitaxial interface with the cubic perovskite lattice, suggesting potential implications for interfacial stability and charge transport [56].

Figure 6a presents Butterworth-filtered LAADF images of the interface between the perovskite FA$_{0.9}$Cs$_{0.1}$PbI$_{3-x}$Cl$_x$ and trigonal PbI$_2$. The corresponding FT in Figure 6b confirms that the perovskite exhibits a cubic structure along the ⟨001⟩ zone axis, while PbI$_2$ adopts a trigonal structure along the ⟨841⟩ zone axis. The two structures have overlapping reflections, as expected from the nominally coherent structure (and shown in the simulated diffraction patterns in Figures 6d–f). The signal from PbI$_2$ is extremely weak as the area of PbI$_2$ is relatively small, so the pattern is dominated by the perovskite lattice reflection.

While there is a localised region with a coherent interface (yellow rectangle in Figure 6a), closer examination of Figure 6a and the atomic structure at the interface between the cubic perovskite FA$_{0.9}$Cs$_{0.1}$PbI$_{3-x}$Cl$_x$ and trigonal PbI$_2$, shows that the two phases do not in fact meet coherently but there are edge dislocations associated with the presence of the PbI$_2$. This is evident when comparing the crystal lattice of the perovskite phase 'above' the PbI$_2$ phase with that 'below', as indicated by the pink lines. In such cases, the trigonal PbI$_2$ is likely to be detrimental to performance.

The lack of perfect coherence between trigonal PbI$_2$ and the perovskite lattice is highlighted by the GPA strain maps derived from the LAADF-STEM image in Figure 6a, corresponding to $\varepsilon_{xx}$ (strain component along the *x*-direction), $\varepsilon_{yy}$ (strain component along the *y*-direction), and $\varepsilon_{xy}$ (shear strain), are shown in Figures 6e, g, and f, respectively. These maps reveal substantial in-plane strain in both x and y directions, as well as shear strain, at the interface between the cubic perovskite and trigonal PbI$_2$ phases, indicating significant lattice mismatch and interfacial stress, particularly in the region of the edge dislocation.

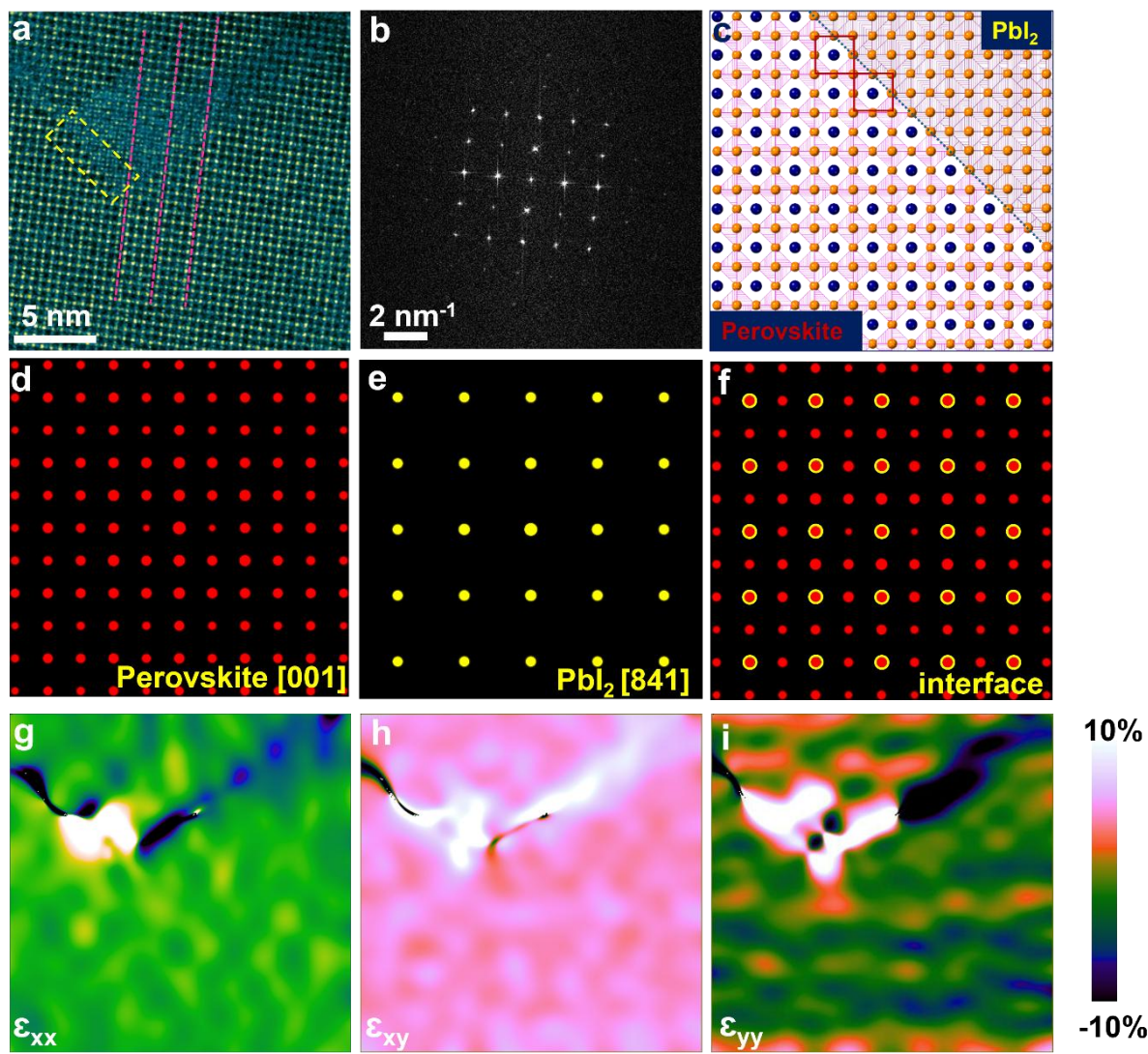

**Figure 6.** The interface between cubic perovskite FA$_{0.9}$Cs$_{0.1}$PbI$_{3-x}$Cl$_x$ and trigonal PbI$_2$. (a) Low-dose atomic-resolution LAADF-STEM image showing the trigonal PbI$_2$ domain within the FA$_{0.9}$Cs$_{0.1}$PbI$_{3-x}$Cl$_x$ lattice. The image is filtered with a Butterworth filter to enhance contrast. The pink lines mark the location of some of the edge dislocations across the interface. (b) FTs of the low-dose LAADF-STEM image shown in (a). (c) Crystal structure model if the interface is coherent between cubic FA$_{0.9}$Cs$_{0.1}$PbI$_{3-x}$Cl$_x$ along ⟨001⟩ zone axis and trigonal PbI$_2$ along ⟨841⟩ zone axis – this model is consistent with the yellow rectangle in Figure a, which is locally dislocation free. (d) Simulated electron diffraction pattern of cubic perovskite along ⟨001⟩ zone axis. (e) Simulated electron diffraction pattern of trigonal PbI$_2$ along ⟨841⟩ zone axis. (f) Simulated electron diffraction pattern of interface between cubic FA$_{0.9}$Cs$_{0.1}$PbI$_{3-x}$Cl$_x$ along ⟨001⟩ zone axis and trigonal PbI$_2$ along ⟨841⟩ zone axis. (g-i) Strain field by GPA analysis shows $\varepsilon_{xx}$ along horizontal direction; $\varepsilon_{yy}$ along vertical direction and shear strain $\varepsilon_{xy}$ generated using LAADF-STEM image.

## 3. Discussion and Conclusions

Introduction of the templating interlayer in co-evaporated thin films of $FA_{0.9}Cs_{0.1}PbI_{3-x}Cl_x$ enables the formation of highly oriented co-evaporated perovskite films, with the majority of grains aligned along the ⟨001⟩ zone axis, perpendicular to the substrate. This controlled crystallographic orientation leads to a structurally well-defined buried interface and correlates directly with improved optoelectronic properties and device performance [35].

While well-aligned along ⟨100⟩, the grains are rotated arbitrarily about this axis, yielding a range of GB interfaces. Measuring and understanding the atomic structure of these GBs, including both low-angle and high-angle configurations, is critical for elucidating their impact on optical, optoelectronic properties and device performance.

We find that very low-angle GBs generate several types of dislocation arrays that preserve partial crystallographic continuity but induce significant local strain, with interaction between the strain fields of adjacent dislocations occurring along the array. Strain mapping proves that the strain components transition from positive to negative at the core, or vice versa. These dislocations and associated strain are likely to impact local bandgap and introduce trap states.

High-angle GBs exhibit pronounced structural disorder, with a higher density of point defects, and dangling bonds, which are expected to increase trap densities and promote non-radiative recombination.

In addition to GBs with arbitrary angles, we also observe higher symmetry GBs. Coincident site lattice Σ5 GBs were observed with a well-formed structural interface. While zig-zag arrangements of 90-degree planar GBs (or 'pseudo-twin' boundaries), parallel to the {110} plane were also observed. They have edge-sharing octahedra and enforce stacking faults parallel to the {100} plane with shifts of the [$PbI_6$] octahedra across the GB. These faults again induce significant local strain fields likely to be detrimental to performance.

Finally, we find a very small volume of the co-evaporated film contains $PbI_2$ contaminant phases (corresponding to ~0.22% of diffraction patterns). This phase is predominantly hexagonal with some trigonal grains. While the crystal lattice of trigonal $PbI_2$ and the cubic perovskite are nominally coherent [56] in certain orientations, atomic-resolution images of trigonal $PbI_2$ grains reveal many dislocations at the perovskite/$PbI_2$ interface accompanied by significant local strain, with only small regions of the interface being coherent. This suggests the presence of trigonal $PbI_2$ grains is not necessarily benign.

While all of the GBs and associated defect structures observed here are likely damaging to photophysical properties, their impact may be mitigated slightly by the fact that they do not lie perpendicular to, and hence do not obstruct directly, the path of the current between the transport layers.

The integration of atomic-scale structural characterization with strain mapping establishes a direct structure–strain correlation framework. This approach provides a foundation for systematically probing the electronic consequences of GBs and their associated defects in OIMHPs. Our observations here suggest that all the GBs observed in these films are likely to be harmful to some extent to photophysical properties and further motivates the need to identify strategies to grow single-crystal films. The results of this study provide valuable insights that can guide the growth of high-quality, low-defect OIMHP films, thereby improving the performance of PSC devices.

## Acknowledgements

This work was supported by the Australian Research Council grants Discovery Project DP200103070 and ARC Laureate Fellowship FL220100202 as well as the EPSRC (UK) via EP/T025077/1 and EP/X038777/1. The authors acknowledge use of facilities within the Monash Centre for Electron Microscopy, a node of Microscopy Australia. The FEI Titan3 80-300 FEG-TEM was funded by ARC grant LE0454166 and the Thermo Fisher Scientific Spectra Phi FEG-TEM was funded by ARC grant LE170100118.

## Competing interests

The authors declare no competing interests.

## Supporting Information

Supporting Information is available.